\documentclass[12pt]{iopart}
\usepackage{iopams,bm}
\usepackage{graphicx,cite}

\begin{document}

\title{Spontaneous emission and teleportation in cavity QED}

\author{P P Munhoz\footnote[1]{Corresponding author: pmunhoz@ifi.unicamp.br}, J A Roversi, A Vidiella-Barranco and F L Semi\~ao}

\address{Instituto de F\'\i sica ``Gleb Wataghin'' - Universidade Estadual de
Campinas, 13083-970 Campinas S\~ao Paulo Brazil}

\begin{abstract}
In this work, we consider atomic spontaneous emission in a system
consisting of two identical two-level atoms interacting
dispersively with the quantized electromagnetic field in a high-Q
cavity. We investigate the destructive effect of the atomic decay
on the generation of maximally entangled states, following the
proposal by Zheng S B and Guo G C (2000 {\it Phys. Rev. Lett.}
{\bf 85} 2392). In particular, we analyze the fidelity of
teleportation performed using such a noisy channel and calculate
the maximum spontaneous decay rate we may have in order to realize
teleportation.
\end{abstract}




\section{Introduction}

The generation and coherent control of maximally entangled states
is of fundamental importance in the achievement of various
information processing tasks as superdense coding \cite{bennett1}
and quantum teleportation \cite{bennett2}. If a pure but not
maximally entangled state is used to perform the standard
teleportation scheme one faces two possibilities: either it is
possible to teleport only particular qubit states, or
teleportation is done with reduced fidelity. Some clever ideas
intending to increase such a teleportation fidelity have been
reported. One possibility is to concentrate partial entanglement
by using local operations and classical communication
\cite{bennett3}. In this case, the sender (Alice) and the receiver
(Bob) are supplied with $n$ pairs of identical partial entangled
states and, for large $n$, the fraction of maximally entangled
states obtained is $nE$, being $E$ the von Neumann entropy of the
density matrices obtained by tracing out one of the subsystems.
Another interesting possibility is the development of optimal
strategies that slightly change the standard protocol to maximize
the probability of perfect teleportation \cite{guo1}. In a more
realistic situation, however, the state shared by Alice and Bob
will happen to be mixed rather than pure due to unwanted noise and
imperfections. In this case, it is still possible to distill out
some quantum entanglement from those states. If the fidelity of
the mixed state relative to a perfect singlet is larger than
$\case12$ and Alice and Bob are provided with many identical
states, there is still a purification scheme that yields some
small fraction of almost perfectly pure singlet states that can be
used to high fidelity quantum teleportation \cite{bennett4}.

The practical implementation of teleportation requires a high degree
of control upon physical systems. The first experimental
teleportation reported made use of correlated photon pairs produced
by parametric down-conversion \cite{zeilenger1}. Since then, it has
been experimentally carried out in a wide variety of quantum systems
\cite{teleport}. In cavity quantum electrodynamics (QED), one of the
most interesting proposals is based on the dispersive interaction of
two identical atoms with a single-mode cavity field (cavity-assisted
atomic collisions) \cite{zheng}. In the ideal case, this process
would lead to the generation of perfectly entangled states
(noiseless channel). From the experimental side, although
teleportation using this system has not yet been performed, the
coherent coupling between the atoms has already been demonstrated
\cite{haroche}. Experimental imperfections (three-body collisions,
for instance) cause the fidelity and entanglement of the generated
state to decrease and this destructive effect can indirectly be seen
in experimentally accessible joint detection probabilities. In the
experiment in reference \cite{haroche}, two Rydberg atoms cross a
nonresonant cavity and become entangled by virtual emission and
absorption of microwave photons. As a consequence, the procedure is
supposed to be insensitive to thermal photons and cavity decay.
However, one may also think of possible decoherence channels for the
atoms, such as atomic spontaneous emission, as having destructive
effects. If one of the atoms decays, the global atomic state
factorizes and all the entanglement is lost. In this paper, we
include spontaneous emission in the two-atom system interacting
through a cavity field and analyze the problem of generation of
maximally entangled states as well as its use in teleportation. We
compute the joint detection probabilities and analyze their
dependence on the atomic decaying rate for different atom-field
detunings. Once the spontaneous emission leads the two-atom system
to a mixed state, we find one of the main results of our paper: an
upper limit for the value of the spontaneous emission rate below
which one can perform quantum teleportation using that particular
noisy channel. We would like to point out that although dissipation
has normally destructive effects on quantum coherence, it may also
allow the generation of entangled atomic states \cite{plenio} and
nonclassical states of the vibrational motion of a trapped ion
\cite{semiao}. In other studies it is shown that an adequate driving
of the system may compensate losses, yielding the stabilization of
entanglement between two atoms \cite{cakir}.

\section{Solution of the Master Equation}

\begin{figure}[t]
\centering
\includegraphics[width=0.7\columnwidth]{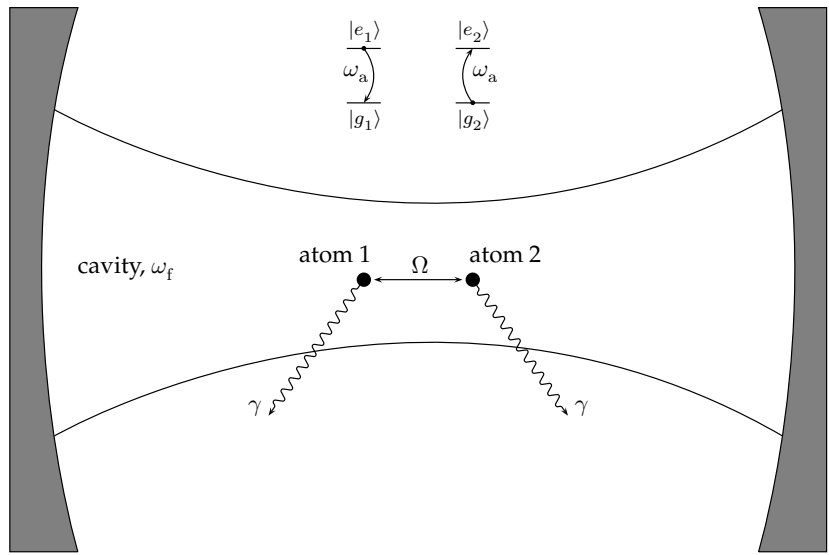}
\caption{\label{fig1}A sketch of the system comprising a couple of two-level atoms
(frequencies $\omega^{}_{\rm a}$), in a far detuned high-Q cavity
(frequency $\omega^{}_{\rm f}$). The atoms interact with each other
via the effective cavity-assisted atom-atom coupling $\Omega$. We assume
that the decay rate $\gamma$ is the same for each atom.}
\end{figure}

We consider a system consisting of two identical two-level atoms
interacting with a quantized electromagnetic field enclosed in a
high-$Q$ cavity \cite{twoatoms}, as shown in figure (\ref{fig1}). If
the two atoms are sufficiently far from each other, we can neglect
the dipole-dipole interaction and the Hamiltonian for this system in
the interaction picture reads \cite{zheng,tavis}
\begin{eqnarray}\label{tcm}
\bm{H}=\hbar\lambda\sum_{j=1,2}(\e^{-\imath\delta{}t}\bm{a}^{\dag}_{}\bm\sigma^{}_{j}+\e^{\imath\delta{}t}\bm{a}^{}_{}\bm\sigma^{\dag}_{j}),
\end{eqnarray}
where $\lambda$ is the atom-cavity coupling constant,
$\bm\sigma^{}_{j}$ ($\bm\sigma^{\dag}_{j}$) is the atomic lowering
(raising) operator, $\bm{a}$ ($\bm{a}^{\dag}_{}$) is the
annihilation (creation) operator of photons in the field mode and
$\delta$ the detuning between the atomic and cavity field
frequencies denoted by $\omega^{}_{\rm a}$ and $\omega^{}_{\rm
f}$, respectively. As considered in the original maximally
entangled state generation proposal \cite{zheng}, we take here the
dispersive limit ($\delta\gg\lambda$) in Hamiltonian (\ref{tcm}).
If there are no photons initially in the cavity (vacuum state),
the system may be described by an effective two-atom Hamiltonian
which, in the interaction picture, is given by \cite{zheng}
\begin{eqnarray}\label{H}
\bm{H}=\hbar\Omega(\bm\sigma^{\dag}_{1}\bm\sigma^{}_{1}+\bm\sigma^{\dag}_{2}\bm\sigma^{}_{2}+
\bm\sigma^{\dag}_{1}\bm\sigma^{}_{2}+\bm\sigma^{}_{1}\bm\sigma^{\dag}_{2}),
\end{eqnarray}
where $\Omega=\lambda^2/\delta$ is the effective cavity-assisted
atom-atom coupling constant.

The above Hamiltonian governs the dynamics of the system in the
ideal case where losses are not included. The Hamiltonian
(\ref{H}) might lead to the generation of a maximally entangled
state. Consider, for instance, that atom $1$ is initially prepared
in the state $|e^{}_{1}\rangle$ and the atom $2$ in the state
$|g^{}_{2}\rangle$, i.e., the initial global state of the two
atoms is
$|\psi(0)\rangle=|e^{}_{1}\rangle\otimes|g^{}_{2}\rangle\equiv|eg\rangle$.
In this case, the evolved state will be \cite{zheng}
\begin{eqnarray}\label{psit}
|\psi(t)\rangle=\e^{-\imath\Omega{}t}[\cos(\Omega{}t)|eg\rangle-\imath\sin(\Omega{}t)|ge\rangle].
\end{eqnarray}
For interaction times\footnote{Or equivalently, detunings
$\delta^{}_{k}=4 \lambda^2t/(2k+1)\pi$.}
$t^{}_{k}=(2k+1)\pi/4\Omega$ ($k$ integer) that corresponds to the
maximally entangled state
\begin{eqnarray}\label{max}
|\psi^{}_{\rm EPR}\rangle=\frac{1}{\sqrt{2}}(|eg\rangle-\imath|ge\rangle),
\end{eqnarray}
which can be used for faithful teleportation as reported in
\cite{zheng}.

However, in a more realistic situation, mechanisms of losses may be
present. Two of the most important dissipation channels are the
leaking of photons through the cavity walls and the atomic
spontaneous emission of photons into noncavity field modes. In
general, losses can be described as the irreversible coupling of the
system to a large reservoir. In the zero temperature case, and
considering the cavity field initially in the vacuum state, the
losses due to cavity damping are irrelevant according to the
effective Hamiltonian (\ref{H}). For this reason, we consider here
just the atomic spontaneous emission which is expected to degrade
the maximum amount of entanglement in state (\ref{max}) until the
system reaches its ground state $|gg\rangle$ (disentangled state).
In the rotating wave and Born-Markov approximations, the density
operator for two atoms obeys the following master equation
\cite{walls,gardiner}
\begin{eqnarray}\label{me}
\dot{\bm\rho}(t)=\frac{1}{\imath\hbar}[\bm{H},\bm\rho(t)]+\bm{{\cal L}}^{}_{\rm a}\bm\rho(t),
\end{eqnarray}
where
\begin{eqnarray}\label{liou}
\bm{{\cal L}}^{}_{\rm a}=2\gamma(\bm\sigma^{}_{1}\cdot\bm\sigma^{\dag}_{1}+\bm\sigma^{}_{2}\cdot\bm\sigma^{\dag}_{2}-\case12\{\bm\sigma^{\dag}_{1}\bm\sigma^{}_{1}+\bm\sigma^{\dag}_{2}\bm\sigma^{}_{2},\cdot\}),
\end{eqnarray}
and $\gamma$ is the spontaneous emission rate\footnote{For
simplicity we assume that both atoms are of the same type and are
placed inside the cavity in positions in which the electric field is
basically the same. Therefore the atoms should have equal
spontaneous emission rates.}. We have used the superoperator
notation $\bm{O}\cdot$ ($\cdot\bm{O}$) which represents the action
of an operator $\bm{O}$ to the left (right) on the target operator
and this convention will be used hereafter. Once we are considering
that the atoms are not close enough to take their mutual interaction
into account, the atom-atom cooperation induced by their coupling
with a common reservoir is not included in master equation
(\ref{me}).

As we show next, one can exactly solve (\ref{me}) by means of the
application of the following unitary transformation
\cite{nicolosi}
\begin{eqnarray}\label{U}
\bm{U}=\exp\left[-\frac{\pi}{4}(\bm\sigma^{\dag}_{1}\bm\sigma^{}_{2}-\bm\sigma^{}_{1}\bm\sigma^{\dag}_{2})\right].
\end{eqnarray}
The operator $\bm{U}$ commutes with
$\bm{N}=\bm\sigma^{z}_{1}+\bm\sigma^{z}_{2}$, i.e.,
$[\bm{U},\bm{N}]=0$, i.e., the transformation above preserves the
total excitation number $\bm{N}$, simplifying the solution of the
problem in cases where the initial excitation number is well
defined. In this work we will consider the case $N=0,1$ because it
is consistent with the initial preparation considered in
\cite{zheng}. Now we may restrict ourselves to the Hilbert
subspace spanned by the basis
$\{|g^{}_{1}g^{}_{2}\rangle,|e^{}_{1}g^{}_{2}\rangle,|g^{}_{1}e^{}_{2}\rangle\}$,
and move to a frame according to
\begin{eqnarray}\label{V}
\bm{V}=\exp\left[\imath\Omega(\bm\sigma^{\dag}_{1}\bm\sigma^{}_{1}+\case{1}{2}\bm\sigma^{z}_{1})t\right].
\end{eqnarray}
We can rewrite the master equation (\ref{me}) as
\begin{eqnarray}\label{meT}
\dot{\widetilde{\bm\rho}}(t)=\bm{T}^{\dag}_{}\dot{\bm\rho}(t)\bm{T}=\bm{{\cal L}}^{}_{\rm a}\widetilde{\bm\rho}(t),
\end{eqnarray}
where $\bm{T}=\bm{U}\bm{V}$ and, after defining the superoperators
\begin{eqnarray}
\begin{array}{lll}
\bm{J}=\bm{J}^{}_{1}+\bm{J}^{}_{2},&\quad\bm{J}^{}_{i}=\bm\sigma^{}_{i}\cdot\bm\sigma^{\dag}_{i},\\
\bm{L}=\bm{L}^{}_{1}+\bm{L}^{}_{2},&\quad\bm{L}^{}_{i}=-\case{1}{2}\{\bm\sigma^{\dag}_{i}\bm\sigma^{}_{i},\cdot\},
\end{array}
\quad(i=1,2)
\end{eqnarray}
the master equation (\ref{meT}) assumes the simple form
\begin{eqnarray}\label{meTsup}
\dot{\widetilde{\bm\rho}}(t)=2\gamma(\bm{J}+\bm{L})\widetilde{\bm\rho}(t).
\end{eqnarray}
Its solution may be written as \cite{mtqo}
\begin{eqnarray}\label{rhoTt}
\widetilde{\bm\rho}(t)=\exp[(\e^{2\gamma{}t}-1)\bm{J}]\exp(2\gamma{}t\bm{L})\widetilde{\bm\rho}(0),
\end{eqnarray}
where $\widetilde{\bm\rho}(0)=\bm{T}^{\dag}_{}\bm\rho(0)\bm{T}$.
Considering the initial state to be
$\bm\rho(0)=|eg\rangle\langle{}eg|=\bm\sigma^{\dag}_{1}\bm\sigma^{}_{1}\bm\sigma^{}_{2}\bm\sigma^{\dag}_{2}$,
we have
\begin{eqnarray}\label{rhoT0}
\widetilde{\bm\rho}(0)=\case{1}{2}(\bm\sigma^{\dag}_{1}\bm\sigma^{}_{1}\bm\sigma^{}_{2}\bm\sigma^{\dag}_{2}-\bm\sigma^{\dag}_{1}\bm\sigma^{}_{2}-\bm\sigma^{}_{1}\bm\sigma^{\dag}_{2}+\bm\sigma^{}_{1}\bm\sigma^{\dag}_{1}\bm\sigma^{\dag}_{2}\bm\sigma^{}_{2}),
\end{eqnarray}
and then, the evolved state according to (\ref{meTsup}) is given
by
\begin{eqnarray}\label{rhoT}
\fl\widetilde{\bm\rho}(t)=\case{1}{2}\e^{-2\gamma{}t}[\bm\sigma^{\dag}_{1}\bm\sigma^{}_{1}\bm\sigma^{}_{2}\bm\sigma^{\dag}_{2}-\bm\sigma^{\dag}_{1}\bm\sigma^{}_{2}-\bm\sigma^{}_{1}\bm\sigma^{\dag}_{2}+\bm\sigma^{}_{1}\bm\sigma^{\dag}_{1}\bm\sigma^{\dag}_{2}\bm\sigma^{}_{2}+2(\e^{2\gamma{}t}-1)\bm\sigma^{}_{1}\bm\sigma^{\dag}_{1}\bm\sigma^{}_{2}\bm\sigma^{\dag}_{2}].
\end{eqnarray}
It is not difficult to show that, when transformed back to the
original space, the state (\ref{rhoT}) reduces to (\ref{psit}) in
the ideal case ($\gamma=0$). From the density operator
$\widetilde{\bm\rho}$, all the statistical properties of the system
may be readily obtained\footnote{One should remember that for a
given observable operator $\bm{A}$, its expectation value
$\langle\bm{A}\rangle$ is given by either $\Tr[\bm\rho\bm{A}]$ or
$\Tr[\widetilde{\bm\rho}\widetilde{\bm{A}}]$, being the latter
referred to any space transformed by unitary operations.}.

\section{Dynamics in the Lossy Case}

Now, we turn to the study of the effect of atomic spontaneous
emission on the system dynamics. We focus on two quantities that
may be readily measured in this system \cite{haroche,haroche2}.

\subsection{Joint Detection Probabilities}

The first quantity is the probability of detecting the two atoms
in one of the four states
$\{|gg\rangle,|ge\rangle\,|eg\rangle,|ee\rangle\}$. Once there is
no pumping mechanism acting upon the system, the state
$|ee\rangle$ remains unpopulated. The joint probabilities for the
initial preparation (\ref{rhoT0}) can be computed from
(\ref{rhoT}) and they are given by
\begin{eqnarray}\label{joints}
&&P(e^{}_{1},g^{}_{2})=\e^{-2\gamma{}t}\cos^2(\Omega{}t),\\
&&P(g^{}_{1},e^{}_{2})=\e^{-2\gamma{}t}\sin^2(\Omega{}t),\\
&&P(g^{}_{1},g^{}_{2})=1-\e^{-2\gamma{}t}.
\end{eqnarray}

\begin{figure}[t]
\centering
\includegraphics[width=0.5\columnwidth]{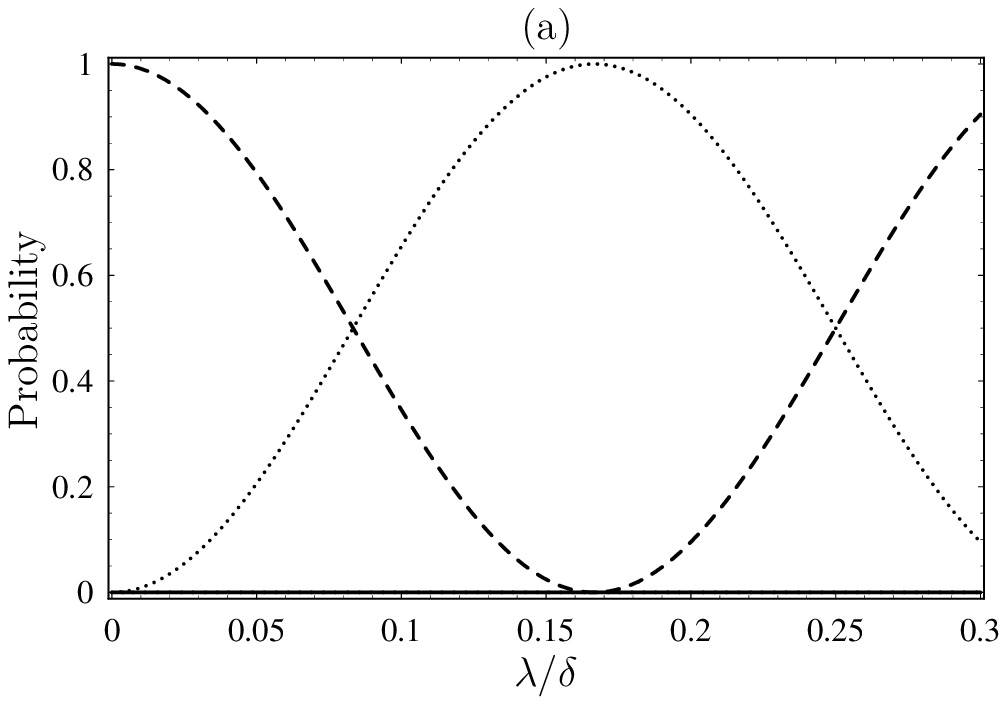}\includegraphics[width=0.5\columnwidth]{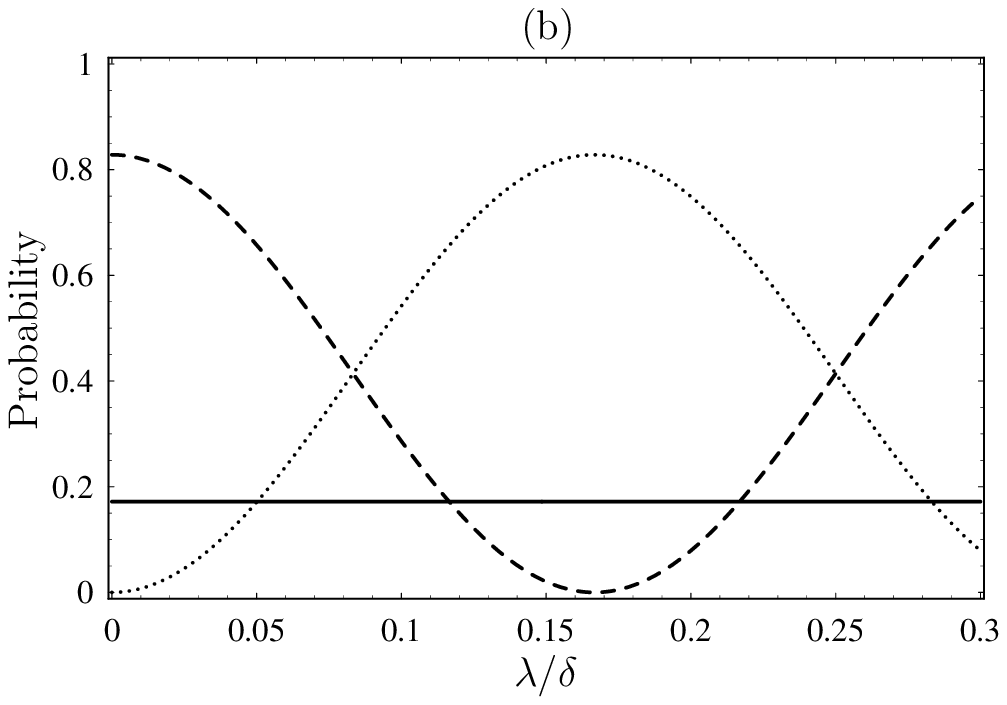}\\
\includegraphics[width=0.5\columnwidth]{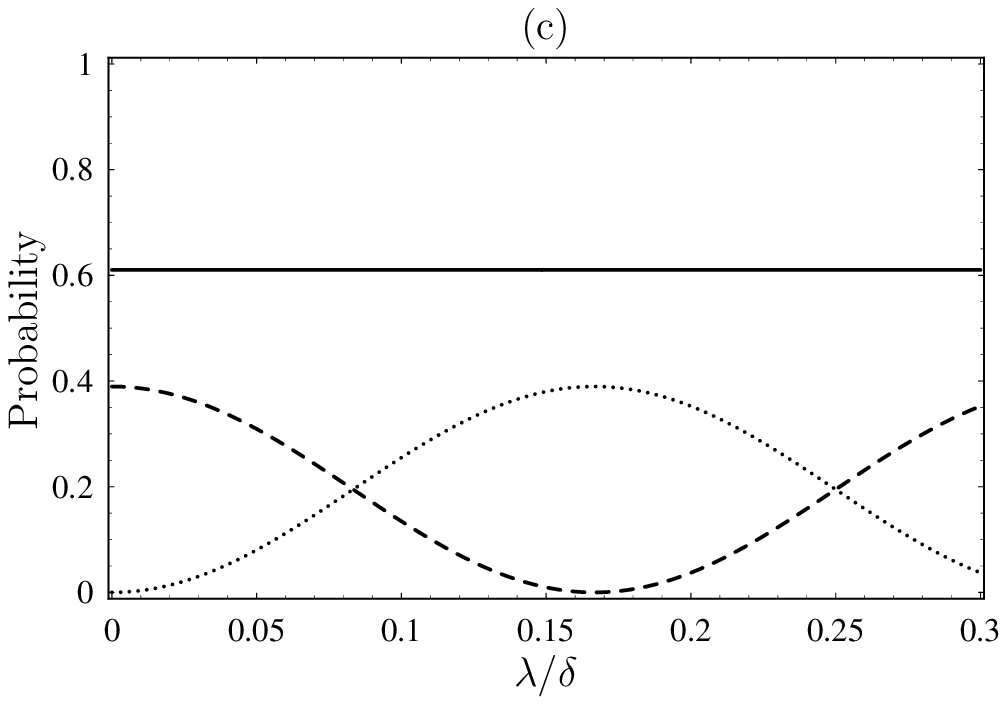}\includegraphics[width=0.5\columnwidth]{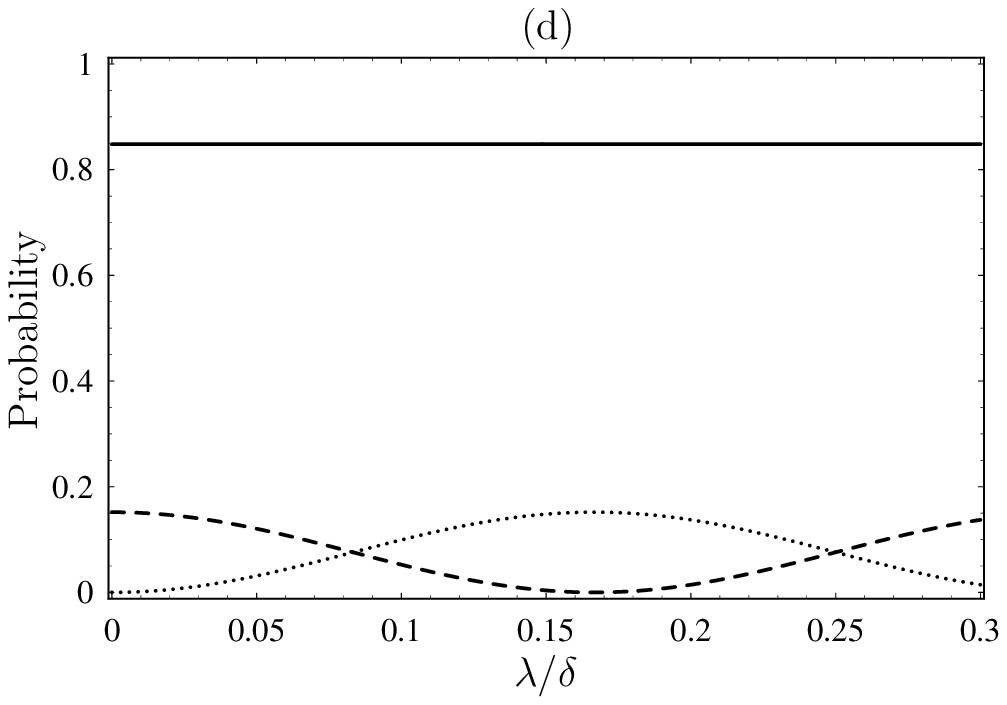}
\caption{\label{fig2}Joint probababilities versus the ratio
$\lambda/\delta$. $P(g^{}_{1},g^{}_{2})$, $P(e^{}_{1},g^{}_{2})$,
and $P(g^{}_{1},e^{}_{2})$ are represented by solid, dashed, and
dotted lines, respectively. For all plots, $\lambda{}t$ is set
equal to $3\pi$. (a) $\gamma/\lambda=0$; (b)
$\gamma/\lambda=0.01$; (c) $\gamma/\lambda=0.05$; (d)
$\gamma/\lambda=0.1$.}
\end{figure}

In figure(\ref{fig2}), we show these probabilities as a function
of the atom-field detuning following the lines in \cite{haroche},
where the experimental curve is presented. In figure \ref{fig2},
one may see that the spontaneous emission plays an important role
in the joint detection probabilities. In the ideal case
($\gamma/\lambda=0$), i.e., the absence of spontaneous emission,
it happens that for times when $P(e^{}_{1},g^{}_{2})$ is equal to
$P(g^{}_{1},e^{}_{2})$, it occurs the generation of the EPR state
(\ref{max}) and neither $|gg\rangle$ nor $|ee\rangle$ is
populated. The situation is quite different when one considers the
more realistic setup where the atoms may decay spontaneously.
Although $P(e^{}_{1},g^{}_{2})$ and $P(g^{}_{1},e^{}_{2})$ still
cross each other, their amplitudes become now suppressed due to
the fact that the stronger the decay rate $\gamma$, the more
important is the mixture with the component $|gg\rangle$. Not
surprisingly, that must be the case, because the system ground
state is expected to become more populated as the atomic levels
couple strongly to the reservoir. When comparing those results to
the experimental curves, one immediately finds common features. As
pointed out in \cite{haroche}, imperfections such as erroneous
detection counts or a third atom crossing the cavity, both could
lead to non null spurious probabilities $P(g^{}_{1},g^{}_{2})$ and
$P(e^{}_{1},e^{}_{2})$ and, consequently, to the suppression of
joint probabilities involving other states \cite{haroche,guo}.
However, according to what it is shown here, even if those
experimental imperfections were fixed, e.g., by allowing just two
atoms inside the cavity at a time and by improving the detection
efficiency, it would still be a spurious population in the ground
state due to atomic spontaneous emission. For the same reason the
quantity $P(e^{}_{1},g^{}_{2})$ would be suppressed in the limit
$\lambda/\delta\rightarrow0$ (infinite detuning). Of course, these
effects become less important as long lived atomic levels are
chosen. It is noteworthy that for small detunings one should use
the original Hamiltonian (\ref{tcm}) rather than the effective
(dispersive) one given by equation (\ref{H}). For the purpose of
the generation of the maximally entangled state, however, one is
interested in the dispersive limit, that starts to be a good
approximation for $\delta>3\lambda$ \cite{haroche}.

\subsection{Bell Signal Analysis}

\begin{figure}[t]
\centering
\includegraphics[width=0.65\columnwidth]{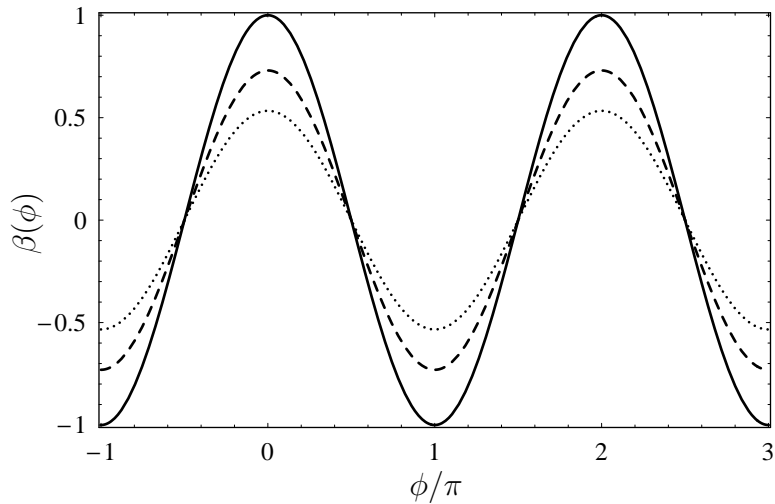}
\caption{\label{fig3}Bell signal as a function of the angular
variable $\phi$. For all plots, $\Omega{}t$ is set equal to
$\pi/4$. We considered the ratio $\gamma/\Omega=0$ (solid),
$\gamma/\Omega=0.2$ (dashed), and $\gamma/\Omega=0.4$ (dotted).}
\end{figure}

The second quantity of interest is the expectation value
$\beta(\phi)\equiv\langle\bm\sigma^{x}_{1}\bm\sigma^{\phi}_{2}\rangle$,
where
$\bm\sigma^{\phi}_{i}=\cos(\phi)\bm\sigma^{x}_{i}+\sin(\phi)\bm\sigma^{y}_{i}$,
being $\bm\sigma^{x}_{i}$ and $\bm\sigma^{y}_{i}$ Pauli matrices for
the atom $i$. Fixing the interaction time to be
$t^{}_{k}=(2k+1)\pi/4\Omega$ ($k$ integer), the variation of the
Bell signal (as a function of $\phi$) measures the angular
correlations between the transverse components associated to the
atoms. Consequently, this quantity gives a measure of the purity of
the generated state. It is easy to show that an equally weighted
mixture of the states $|eg\rangle$ and $|ge\rangle$ would give
$\langle\bm\sigma^{x}_{1}\bm\sigma^{\phi}_{2} \rangle=0$ for all
$\phi$. In this sense, looking at the joint detection probabilities
alone is not enough to decide if a pure EPR state has been
generated. In other words, the joint probabilities give information
about the diagonal elements of the density matrix whereas the Bell
signal furnishes information about its non diagonal elements. This
quantity can be computed directly from (\ref{rhoT}) and the result
is
\begin{eqnarray}
\beta(\phi)=\e^{-\gamma(2k+1)\pi/2\Omega}\cos(\phi).
\end{eqnarray}
Again, the effect of spontaneous emission is to cause a
suppression of the oscillations as shown in figure (\ref{fig3}).
The same behaviour was experimentally observed \cite{haroche}.

\section{Fidelity of Teleportation}

\begin{figure}[t]
\centering
\includegraphics[width=0.7\columnwidth]{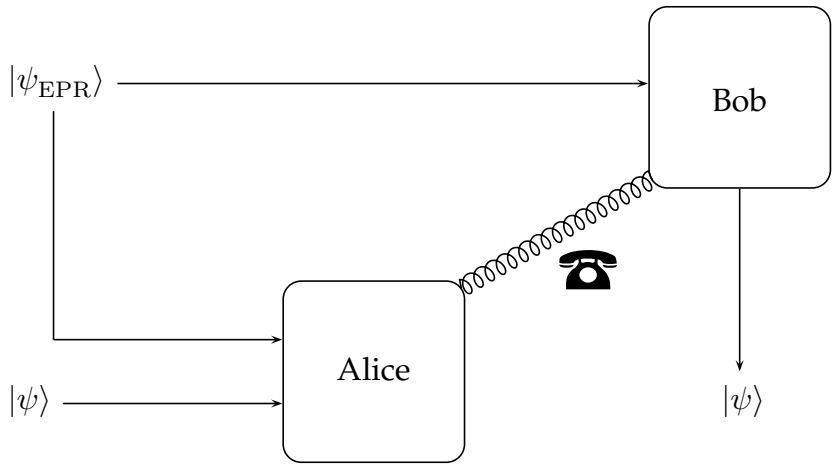}
\caption{\label{fig4}Scheme of quantum teleportation of an unkown
quantum state $|\psi\rangle$. Alice and Bob share a previously
generated quantum channel, a EPR pair. After combining the state
$|\psi\rangle$ with her portion of the EPR pair, Alice performs a
Bell state measurement, destroying the original state. She then
informs Bob the result of her measurement, so that he may perform
the correct operation on his portion of the EPR pair and thus
recover the original teleported state $|\psi\rangle$.}
\end{figure}

In the standard teleportation protocol \cite{bennett2}, two parties
Alice and Bob are supplied with a pair of particles in a EPR state.
They keep their respective part with one another to function as a
quantum channel to teleport a third particle in possession of Alice,
which is in an unkown pure state $|\psi\rangle = a |e\rangle +
b|g\rangle$. Alice then separately performs a joint measurement on
her part of the EPR pair and on the state she wants to teleport, and
sends the result of her measurement to Bob using a classical channel
(e.g., a phone call). As the quantum channel is maximally entangled,
no matter what outcome from the Bell measurement Alice gets, Bob
will have his part of the channel projected into one of four
possible transformations of the original but no longer existing pure
state $|\psi\rangle$. To accomplish the teleportation procedure, Bob
only needs to apply, based on the information received from Alice, a
combination of the four transformations $(\bm{I}, \bm\sigma^{x},
\bm\sigma^{y}, \bm\sigma^{z})$. The protocol is depicted in figure
(\ref{fig4}).

Maximally entangled states provide a quantum channel that allows the
perfect teleportation of an unknown state from one party to
{another. After the seminal paper \cite{bennett2}, it seemed natural
to search for connections between teleportation, Bell's inequalities
and inseparability \cite{popescu1,horodecki1,massar,gisin}. One of
the first things that had been noticed was the fact that a mixed
state could still be useful for (imperfect) teleportation
\cite{popescu1}; it was demonstrated that the fidelity of
transmission of an unknown qubit can be linked to the nonclassical
character of the state forming the quantum channel. Actually, a
classical channel can give at most a fidelity equal to $\case23$.
This may be achieved if Alice simply performs a measurement on the
unkwown state and tells the result to Bob \cite{popescu1}.

\begin{figure}[t]
\centering
\includegraphics[width=0.7\columnwidth]{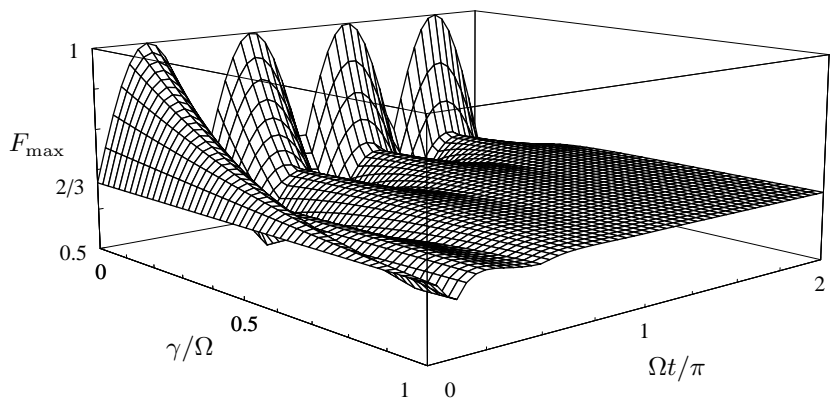}
\caption{\label{fig5}Fidelity of teleportation as a function of the scaled
time $\Omega t$ and decay rate $\gamma/\Omega$.}
\end{figure}

In the system treated here, the two atoms become maximally entangled
at specific interaction times when no imperfections or losses are
considered in the model. Then, a practical question that naturally
arises is: What is the maximum value of the decay constant $\gamma$
that still allows legitimate, although imperfect quantum
teleportation? We answer that question using the fidelity of
teleportation discussed in \cite{horodecki1}. They showed that the
maximum fidelity that can be obtained using a noisy channel
$\bm\rho$ is given by
\begin{eqnarray}
F^{}_{\rm{max}}=\case12(1+\case13\Tr\sqrt{T^\dag{}T}),
\end{eqnarray}
where $T$ is a real matrix formed by the elements
$t^{}_{\rm{nm}}=\rm{Tr}[\bm\rho(\bm\sigma_1^n\otimes\bm\sigma_2^m)]$,
where $\{\bm\sigma_i^n\}^{3}_{\rm{n=1}}$ are the ordered Pauli
matrices $\{\bm\sigma_i^x,\bm\sigma_i^y,\bm\sigma_i^z\}$ for the
atom $i$. In the interaction times $t^{}_{k}$, the maximum fidelity
of teleportation may be obtained from (\ref{rhoT}) and it is given
by
\begin{eqnarray}
F^{}_{\rm{max}}=\cases{
\case13+\case23\e^{-\gamma(2k+1)\pi/2\Omega},&for
$\gamma/\Omega\leqslant\ln(4)/(2k+1)\pi$\\
\case23,&otherwise.}
\end{eqnarray}
For those interaction times, a fidelity of $\case23$ indicates that
the system has decayed to a state that no longer allows reliable
quantum teleportation. The higher the $\gamma$, the closer the
system state gets closer to the ground state $|gg\rangle$. Now, we
can answer the question raised above: the system provides a quantum
channel suitable for teleportation, at interaction times $t^{}_{k}$,
provided the decay constant $\gamma$ does not exceed the upper limit
$\gamma=\Omega\ln(4)/(2k+1)\pi$.

Of course, it is not just for the specific interaction times
$t^{}_{k}$ that one would obtain $F^{}_{\rm{max}}>\case23$.
Precisely at those times the function $F^{}_{\rm{max}}$ has local
maxima. This can be seen in figure (\ref{fig5}), where there is a
plot of $F^{}_{\rm{max}}$ as a function of the scaled time and the
decay rate.}

\section{Entanglement}

\begin{figure}[t]
\centering
\includegraphics[width=0.7\columnwidth]{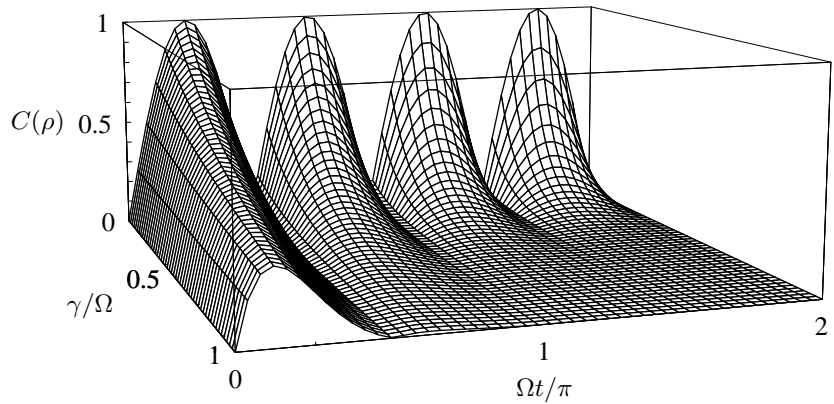}
\caption{\label{fig6}Concurrence as a function of the scaled time $\Omega{}t$
and decay rate $\gamma/\Omega$.}
\end{figure}

In this section we analyze the amount of entanglement between the
two atoms using the concurrence, defined as \cite{wootters}
\begin{eqnarray}\label{conc}
C(\rho)=\max{\{0,\lambda^{}_{1}-\lambda^{}_{2}-\lambda^{}_{3}-\lambda^{}_{4}\}},
\end{eqnarray}
where $\lambda^{}_{i}$ are the square roots of the eigenvalues, in
decreasing order, of the non-Hermitian matrix $\rho\rho^{}_{\rm
sf}$, being
\begin{eqnarray}\label{rhosf}
\rho^{}_{\rm sf}=(\sigma^{y}_{1}\otimes\sigma^{y}_{2})\rho^\ast(\sigma^{y}_{1}\otimes\sigma^{y}_{2}),
\end{eqnarray}
the ``spin flip'' transformation of the complex conjugation of $\rho$ in the standard basis
$\{|ee\rangle,|eg\rangle,|ge\rangle,|gg\rangle\}$, and
$\sigma^{y}_{i}$ is the Pauli spin matrix. For a maximally entangled state $C(\rho) = 1$, while for a
separable state $C(\rho) = 0$.

At the interaction times $t^{}_{k}$, the concurrence is given by
\begin{eqnarray}
C(\rho)=\e^{-\gamma(2k+1)\pi/2\Omega},
\end{eqnarray}
which is related to the maximum fidelity of teleportation as
\begin{eqnarray}
F^{}_{\rm{max}}=\cases{\case13+\case23C(\rho),&for
$\gamma/\Omega\leqslant\ln(4)/(2k+1)\pi$\\
\case23,&otherwise.}
\end{eqnarray}
Therefore, in this case, for $C(\rho)\geqslant\case12$ the system
provides a quantum channel for reliable teleportation. The
dependence of entanglement on time and on the damping rate is shown
in figure (\ref{fig6}), where we have a plot of the concurrence as a
function of $\Omega{}t$ and the decay rate $\gamma/\Omega$. In
comparing figure (\ref{fig6}) with figure (\ref{fig5}) one may find,
for instance, ranges of parameters for which quantum teleportation
is not allowed despite the existence of entanglement.

\section{Conclusions}

We have investigated the effect of atomic spontaneous emission in
the generation of maximally entangled states, which are useful for
quantum teleportation. We have shown that experimental observables
such as joint probabilities and correlation functions involving
transverse components of the Pauli matrices both are suppressed due
to the effect of atomic decay. We have computed the fidelity of
teleportation and found an upper limit for the value of the
spontaneous emission rate below one can perform quantum
teleportation using that noisy channel. The knowledge of the upper
bound for the decay rate allowing the realization of teleportation,
as we have found here, is of central importante if one wants to
implement such a quantum channel in an actual experiment. The effect
considered here may be lessened if the spontaneous emission is set
to be negligible, what would correspond to the choice of long lived
atomic levels. Nevertheless, if one wants to address a more general
situation, the effects of spontaneous emission are not negligible
and have to be taken into account. We have also calculated the
amount of entanglement (quantified by the concurrence) between the
atoms as a function of the atomic decay rate, which shows that a
substantial amount of entanglement is required in order to allow
quantum teleportation using the atomic quantum channel here
discussed.

\ack This work is partially supported by CNPq (Conselho Nacional
para o Desenvolvimento Cient\'\i fico e Tecnol\'ogico), and FAPESP
(Funda\c c\~ao de Amparo \`a Pesquisa do Estado de S\~ao Paulo)
grant number 02/02715-2, Brazil.


\Bibliography{xx}

\bibitem{bennett1}Bennett C H and Wiesner S J 1992 \PRL {\bf 69} 2881

\bibitem{bennett2}Bennett C H \etal 1993 \PRL {\bf 70} 1895

\bibitem{bennett3}Bennett C H \etal 1996 \PR A {\bf 53} 2043

\bibitem{guo1}Li W L \etal 1996 \PR A {\bf 61} 034301

\bibitem{bennett4}Bennett C H \etal 1996 \PRL {\bf 76} 722

\bibitem{zeilenger1}Bouwmeester B \etal 1997 {\it Nature} {\bf 390} 575

\bibitem{teleport}Boschi E \etal 1998 \PRL {\bf 80} 1121

Furusawa A \etal 1998 {\it Science} {\bf 282} 706

Nielsen M A \etal 1998 {\it Nature} {\bf 396} 52

Pan J W \etal 2001 \PRL {\bf 86} 4435

Marcikic I \etal 2002 {\it Nature} {\bf 421} 509

Fattal D \etal 2004 \PRL {\bf 92} 037904

Riebe M \etal 2004 {\it Nature} {\bf 429} 734

Barret M D \etal 2004 {\it Nature} {\bf 429} 737

\bibitem{zheng}Zheng S B and Guo G C 2000 \PRL {\bf 85} 2392

\bibitem{haroche}Osnaghi S \etal 2001 \PRL {\bf 87} 037902

\bibitem{plenio} Plenio M B  \etal 1999 \PR A {\bf 59} 2468

\bibitem{semiao} Semi\~ao F L and Vidiella-Barranco A. 2005 \PR A {\bf 71} 065802

\bibitem{cakir} \c Cakir \"O \etal 2005 \PR A {\bf 71} 034303

\bibitem{twoatoms}Ye L and Guo G C 2005 \PR A {\bf 71} 034304

Xu J B and Li S B 2005 \NJP {\bf 7} 72

\bibitem{tavis}Tavis M and Cummings F W 1968 \PR {\bf 170} 379

Dicke R 1954 \PR {\bf 93} 99

\bibitem{walls}Walls D F and Milburn G J 1994 {\it Quantum Optics} (Berlim: Springer)

\bibitem{gardiner}Gardiner C W and Zoller P 2000 {\it Quantum Noise} (Berlim: Springer)

\bibitem{nicolosi} Nicolosi S \etal 2004 \PR A {\bf 70} 022511

\bibitem{mtqo}Barnett S M and Radmore P M 1997 {\it Methods in Theoretical Quantum Optics} (Oxford: Claredon)

\bibitem{haroche2}Brune M \etal 1996 \PRL {\bf 77} 4887

\bibitem{guo}Guo G P \etal 2002 \PR A {\bf 65} 042102

\bibitem{popescu1}Popescu S 1994 \PRL {\bf 72} 797

\bibitem{horodecki1}Horodecki R \etal 1996 \PL A {\bf 222} 21

\bibitem{massar}Massar S and Popescu S 1995 \PRL {\bf 74} 1259

\bibitem{gisin}Gisin N 1996 \PL A {\bf 210} 157

Gisin N 1996 \PL A {\bf 210} 151

\bibitem{wootters}Wootters W K 1998 \PRL {\bf 80} 2245

\endbib
\end{document}